\documentclass[11pt,a4paper]{iopart}

\begin{document}

\jl{6}

\title{Unified Dark Energy--Dark Matter model with Inverse Quintessence}

\author{Stefano Ansoldi$^{1,2,3}$
and
Eduardo I. Guendelman$^{4}$}

\address{$^{1}$ICRA, International Center for Relativistic Astrophysics}
\address{$^{2}$INFN, Istituto Nazionale di Fisica Nucleare}
\address{$^{3}$Dipartimento di Matematica e Informatica, Universit\`{a} degli Studi di Udine, via delle Scienze 206, I-33100 Udine (UD), Italy}
\address{$^{4}$Department of Physics, Ben-Gurion University of the Negeev, Beer-Sheva 84105, Israel}

\eads{\mailto{ansoldi@fulbrightmail.org}, \mailto{guendel@bgu.ac.il}}

\begin{abstract}
We consider a model where both dark energy and dark matter originate
from the coupling of a scalar field with a non-conventional kinetic term to, both, a metric
measure and a non-metric measure. An interacting dark energy/dark matter scenario can be obtained by introducing
an additional scalar that can produce non constant vacuum energy and associated variations in
dark matter. The phenomenology is most interesting when the kinetic term of the additional scalar field
is ghost-type, since in this case the dark energy vanishes in the early universe and then grows with time.
This constitutes an ``inverse quintessence scenario'', where the universe starts from a zero
vacuum energy density state, instead of approaching it in the future.
\end{abstract}

%\keywords{}

%\arxivnumber{}

\maketitle

\section{Introduction}

The unification of dark matter (DM, for recent reviews see, e.g., \cite{bib:28,bib:29,bib:30,bib:31}) and dark energy (DE, see, for instance, \cite{bib:25,bib:26,bib:27}) using a scalar field $\phi$ with a
non-conventional kinetic term of the form\footnote{The following conventions apply: Greek indices
are used to denote the components of tensors with respect to a given choice of spacetime coordinates (e.g., $g _{\mu \nu}$, $\mu, \nu = 0$, $1$, $2$, $3$, are the components of the metric tensor, which is taken with signature $(-,+,+,+)$, and has determinant $g$); Latin indices $i$, $j$, $k$, $l$, \dots{}, take the values $1$, $2$, $3$, $4$ in the main text and label a set of four scalar fields, defined hereafter; in the appendix the range of variation of these kind of indices is $1$, $2$ when dealing with the string case and $1$, \dots , $p$ when discussing the $p$-brane case. $R$ is the Ricci scalar, considered in the standard general relativistic framework as a function of the metric and its derivatives. The alternating symbol, in a number of dimensions equal to the number of indices of the symbol, is always represented by $\epsilon$ (for instance, $\epsilon ^{\mu \nu \rho \sigma}$ is a quantity that vanishes if at least two indices take the same value, is $+1$ if the indices are an even permutation of $(0,1,2,3)$ and is $-1$ if the indices are an odd permutation of $(0\,1\,2\,3)$). $G$ is Newton gravitational constant, and units are chosen so that the speed of light is equal to one.}
$\sqrt{- \det(g _{\mu\nu} + \partial _{\mu} \phi \partial _{\nu} \phi)}$
was considered in Ref.~\cite{bib:01}. These types of kinetic terms appear, for example, for the
tachyon in string theory. In Ref.~\cite{bib:01} also an arbitrary potential $V (\phi)$ was considered
and an action of the \emph{two measures type} \cite{bib:05,bib:06,bib:07,bib:08,bib:09,bib:10,bib:11,bib:12,bib:13,bib:14,bib:15,bib:32}
\begin{equation}
    S [ g, \varphi ^{i}, \phi]
    =
    \int {\mathcal{L}} _{1} \sqrt{-g} d ^{4} x
    +
    \int {\mathcal{L}} _{2} \Phi d ^{4} x
\label{eq:act01a}
\end{equation}
was analyzed. In the particular case at hand, defined by the specific choice of Ref.~\cite{bib:01}, we take ${\mathcal{L}} _{1}$ and ${\mathcal{L}} _{2}$
as
\begin{equation}
    {\mathcal{L}} _{1} = - \frac{1}{\kappa} R + X
    , \quad
    {\mathcal{L}} _{2} = X
    , \quad
    \kappa = ( 16 \pi G ) ^{-1}
    ,
\label{eq:act01b}
\end{equation}
where
\begin{equation}
    X
    =
    V (\phi)
    \frac{\sqrt{- \det(g _{\mu \nu} + \partial _{\mu} \phi \partial _{\nu} \phi)}}
         {\sqrt{- g}}
\label{eq:act01c}
\end{equation}
and the \emph{density} $\Phi$ is defined by
\begin{equation}
    \Phi
    =
    \epsilon ^{\mu \nu \rho \sigma}
    \epsilon _{i j k l}
    \partial _{\mu} \varphi ^{i}
    \partial _{\nu} \varphi ^{j}
    \partial _{\rho} \varphi ^{k}
    \partial _{\sigma} \varphi ^{l}
    ,
\label{eq:act01d}
\end{equation}
in which $\varphi ^{i}$, $i = 1$, $2$, $3$, $4$ are $4$-scalar fields.

Here we take ${\mathcal{L} _{1}}$ and ${\mathcal{L} _{2}}$ to be $\varphi ^{i}$ independent. The action $S$ can then be written as
\begin{equation}
    \fl
    S [ g, \varphi ^{i}, \phi]
    =
    -
    \int \frac{R}{\kappa} \sqrt{-g} d ^{4} x
    +
    \int V (\phi) \sqrt{- \det(g _{\mu\nu} + \partial _{\mu} \phi \partial _{\nu} \phi)} d ^{4} x
    +
    \int X \Phi d ^{4} x
    ,
\label{eq:expact}
\end{equation}
from which it is readily seen that, apart from the new $\Phi$-dependent term, it coincides with the
tachyon model \cite{bib:02,bib:03,bib:04}.
A similar type of non metric structure like $\Phi$ was used in \cite{bib:13,bib:14,bib:15} to study supergravity models.

The introduction of the new measure $\Phi$ provides new possibilities to realize scale invariance,
since $\Phi$ and $\sqrt{-g}$ could transform differently under scale transformations \cite{bib:08,bib:09}; in
addition one may use such models to construct brane-world scenarios where naturally no four dimensional cosmological
constant is generated and only the extra dimensions get curved \cite{bib:12}. Non singular ``emergent type''
cosmologies are also possible for theories of this type \cite{bib:13,bib:14,bib:15}.

\section{A unified non interacting DE/DM model}

We now briefly review the model defined by eqs.~(\ref{eq:act01a})--(\ref{eq:act01d}).
The variation of the action with respect to the measure fields $\varphi ^{i}$ leads to the equations of motion
\begin{equation}
    A ^{\mu} _{i} \partial _{\mu} {\mathcal{L}} _{2} = 0
    ,
\label{eq:varphieq1}
\end{equation}
where
\begin{equation}
    A ^{\mu} _{i}
    =
    \epsilon ^{\mu \nu \rho \sigma}
    \epsilon _{i j k l}
    \partial _{\nu} \varphi ^{j}
    \partial _{\rho} \varphi ^{k}
    \partial _{\sigma} \varphi ^{l}
    ;
\label{eq:varphieq2}
\end{equation}
since $\det(A ^{\mu} _{i})$ is, up to a constant, equal to $\Phi ^{3}$, for non degenerate measures,
i.e. $\Phi \neq 0$, we have that (\ref{eq:varphieq1}) and (\ref{eq:varphieq2}) imply
\begin{equation}
    {\mathcal{L}} _{2} = M
    ,
\label{eq:L_2con}
\end{equation}
with $M$ constant. For the case of eqs.~(\ref{eq:act01b}) and (\ref{eq:act01c}), in a cosmological setting and with a choice of coordinates such that $g _{0a} = 0$,
$a = 1$, $2$, $3$, $g _{00} < 0$ and $\phi = \phi (t)$, the result (\ref{eq:L_2con}) implies
\[
    X = V (\phi) \sqrt{1 + \frac{\dot{\phi} ^{2}}{g _{00}}} = M = \mathrm{const.}
\]
After squaring the above relation under the condition $\mathrm{sgn} (M V (\phi)) > 0$, we can recast it into
a form that looks like an energy balance equation,
\[
    - \frac{\dot{\phi} ^{2}}{g _{00}} + \frac{M ^{2}}{V ^{2} (\phi)} = 1 ;
\]
indeed, since we have $g _{00} < 0$, the first term on the left hand side looks like a standard (i.e. positive definite) kinetic energy term, while an effective potential (that behaves as the squared inverse of the potential $V (\phi)$) can be read out of the following term and can be stable even if the original $V (\phi)$ is unstable.

We can, then, consider the variation of the action (\ref{eq:expact}) with respect to $g ^{\mu \nu}$, which gives, with standard notation,
\[
    R _{\mu \nu} - \frac{1}{2} g _{\mu \nu} R = T ^{\Phi} _{\mu \nu} + T ^{\phi} _{\mu \nu}
    ,
\]
where $T ^{\Phi} _{\mu \nu}$ is the contribution to the stress-energy tensor coming from the new measure and $T ^{\phi} _{\mu \nu}$ are the usual terms coming from the standard tachyon action.
In the framework of a Friedmann-Robertson-Walker cosmological scenario, we specialize the metric to the form
\[
    ds ^{2} = g _{00} dt ^{2} + a ^{2} (t) \left [ \frac{dr ^{2}}{1 - k r ^{2}} + r ^{2} (d \theta ^{2} + \sin ^{2} \theta d \eta ^{2} )\right ] .
\]
Again $(t,r,\theta,\eta)$ is the standard\footnote{The name for the azimuthal angle $\eta$ is \emph{not standard}, but we are already using the standard choice to denote the scalar field with non-standard kinetic term; in any case, the azimuthal angle will play no role in the following and will not appear anywhere else.} choice for the coordinates commonly used in the literature, where $t$ is the cosmic time,
$k = -1, 0, +1$ represents the constant curvature of the spatial sections $t = \mathrm{const.}$, and $a (t)$ is the scale factor.
The first contribution on the right hand side of the equations obtained from the variation with respect to $g ^{\mu \nu}$ is
different from zero only for $\mu = \nu = 0$. Then the only non vanishing component of $T ^{\Phi} _{\mu \nu}$ can be written as
$T ^{\Phi} _{00} = g _{00} \rho _{\Phi}$, where
\begin{equation}
    \rho _{\Phi} = \frac{\Phi V ^{2} (\phi)}{M \sqrt{-g}} \left( \frac{\dot{\phi} ^{2}}{- g _{00}} \right)
    .
\label{eq:rhoPhi}
\end{equation}
The non vanishing components of the stress-energy tensor contribution from $\phi$ are instead $T ^{\phi} _{00} = g _{00} \rho _{\phi}$ and
$T ^{\phi} _{ij} = - g _{ij} p _{\phi}$, where
\[
    \rho _{\phi} = \frac{V ^{2} (\phi)}{M} , \quad \mathrm{and} \quad p _{\phi} = - M .
\]
Correspondingly, the total energy density and pressure are given by
\[
    \rho _{\mathrm{total}} = \frac{\Phi V ^{2} (\phi)}{M \sqrt{-g}} \left( \frac{\dot{\phi} ^{2}}{- g _{00}} \right) + \frac{V ^{2} (\phi)}{M}
    \quad \mathrm{and} \quad
    p _{\mathrm{total}} = - M .
\]
The total pressure has a rather immediate expression, but to find the total energy density explicitly, we must solve for $\Phi$ first.
We are going to do this in the gauge $g _{00} = -1$, using then the field equations resulting from the action
\[
    S [ g , \varphi ^{i}, \phi]
    =
    \int d ^{4} x
        \left [
            \frac{\sqrt{-g} R}{16 \pi G} + \Phi X + \sqrt{-g} V (\phi) \sqrt{1 - \dot{\phi} ^{2}}
        \right ]
    ,
\]
which, in view of the relation $d / (d t) = \dot{\phi} \, d / (d \phi)$ and defining $\psi = \Phi + \sqrt{-g}$, can be recast into the form
\[
    \frac{d}{d \phi} \left ( \frac{\psi V ^{2}}{M} \sqrt{1 - \frac{M ^{2}}{V ^{2}} }\right)
    =
    - \left( \frac{d V}{d \phi} \right) \frac{\psi M}{V} \left( 1 - \frac{M ^{2}}{V ^{2}} \right) ^{-1/2}
    .
\]
The above equation can be simplified into
\[
    \int \frac{d \psi}{\psi} = - 2 \int \left( 1 - \frac{M ^{2}}{V ^{2}} \right) ^{-1/2} \frac{d V}{V} ,
\]
so that
\[
    \psi = \frac{C (r , \theta)}{V ^{2} - M ^{2}}
\]
or, going back to $\Phi$,
\begin{equation}
    \Phi = \frac{C (r , \theta)}{V ^{2} - M ^{2}} - \sqrt{-g} .
\label{eq:PhiMetMas}
\end{equation}
The undetermined function $C (r , \theta)$, which is \emph{time independent}, can be fixed as
$C (r, \theta) = K r ^{2} \sin \theta (1 - k r ^{2}) ^{-1/2}$, where $K$ is a constant. This choice for $C (r, \theta)$ leads to an
energy density that is independent of the spatial coordinates, as the dependence of $\sqrt{-g}$ from them is exactly cancelled by
the dependence of $C (r , \theta)$. Substituting for $\Phi$ in the expression for the total energy density, we recognize two
additive contributions, a constant one (typical of a cosmological constant term) and an $a ^{-3}$ dependent one, typical
of dust. Indeed
\[
    \rho _{\mathrm{total}} = M + \frac{K}{M a ^{3}} .
\]

\section{Unified dynamical DE and DM}

As we have seen in the previous section the constant of integration $M$ serves as a ``floating'' vacuum energy
or ``floating'' DE, since $M$ is an undetermined constant.

We now would like to promote this DE to a dynamical variable. In this respect, we should recall the analogous
situation when formulating string and brane theories with a dynamical tension, but extending the study to the case
in which there is a non metric measure $\Phi$ in the world-sheet of the brane or string \cite{bib:16,bib:17,bib:18,bib:19,bib:20,bib:21,bib:22}.
The relevant background is detailed in appendix~\ref{app:dynten}. Here we, instead, proceed to generalize those ideas
developed for string and brane tensions, and apply them to DE, that will then become dynamical. Moreover,
since DE and DM will be studied in a unified scheme, the DM will also turn out to be
dynamical. For this we introduce, as we did in the case of strings and branes, a new scalar field $\vartheta$ and
consider the full action, involving $\phi$, $g _{\mu \nu}$, $\varphi ^{i}$ (through the measure $\Phi$) and, now,
$\vartheta$, with the following form:
\[
    S  [ g, \varphi ^{i}, \phi, \vartheta]
    =
    \int {\mathcal{L}} _{1} \sqrt{-g} d ^{4} x
    +
    \int {\mathcal{L}} _{2} \Phi d ^{4} x
    ,
\]
where now
\[
    {\mathcal{L}} _{1} = \frac{1}{\kappa} R + X + {\mathcal{L}} _{(\vartheta)}
\]
and
\[
    {\mathcal{L}} _{2} = X + f \vartheta \; , \quad \mbox{where $f$ is some coupling costant.}
\]
As before, we still have
\begin{enumerate}
\item[i)] $X$ defined by (\ref{eq:act01c}) and
\item[ii)] $\Phi$ being the non metric measure given by (\ref{eq:act01d}).
\end{enumerate}
Our choice for ${\mathcal{L}} _{(\vartheta)}$ will be, instead, discussed later on. Now, unlike the
case when the scalar field $\vartheta$ was absent, the presence of a non trivial $V (\phi)$ makes the problem untractable
analytically in full generality. For this reason, in this paper, we will start concentrating on the case in which
$V (\phi) = V$ is a real constant. Under this condition, the equation for $\phi$ simplifies considerably, so that it allows
an exact solution. Even in this simplified scheme, the non trivial DE and DM dynamics is quite interesting as we will see.

To see this we start with the equation obtained from the variation of the $\varphi ^{i}$ scalars, namely
\[
    A ^{\mu} _{i} \partial _{\mu} \left( X + f \vartheta  \right) = 0 ,
\]
which can be solved to get, for non-degenerate $\Phi$,
\[
    X + f \vartheta = M ,
\]
where again $M$ is a constant. In a cosmological scenario, choosing coordinates as above, having then again the components of the metric such that $g _{0a} = 0$, and with $\phi = \phi (t)$, $\vartheta = \vartheta (t)$, we obtain
\[
    X = V \sqrt{1 + \frac{\dot{\phi} ^{2}}{g _{00}}} = M - f \vartheta
\]
that gives
\[
    \frac{(M - f \vartheta) ^{2}}{V ^{2}} - \frac{\dot{\phi} ^{2}}{g _{00}} = 1
    .
\]
In the gauge $g _{00} = -1$, we can rewrite the above as
\begin{equation}
    \sqrt{1 - \dot{\phi} ^{2}} = \frac{M - f \vartheta}{V}
    , \quad \mathrm{from\ which} \quad
    \dot{\phi} = \sqrt{1 - \frac{(M - f \vartheta) ^{2}}{V ^{2}}}
    .
\label{eq:phiphidot}
\end{equation}
For $V = \mathrm{const.}$, the equation of motion for $\phi$ simplifies to\footnote{We also set $g _{00} = -1$ as before.}
\[
     \frac{d}{dt}
     \left(
        \frac{V \Phi \dot{\phi}}{\sqrt{1 - \dot{\phi} ^{2}}}
        +
        \frac{V \sqrt{-g} \dot{\phi}}{\sqrt{1 - \dot{\phi} ^{2}}}
     \right)
     =
     0
     .
\]
Defining $\psi = \Phi + \sqrt{-g}$, from the equation just written we obtain\footnote{Please, note that the arbitrary function $C (r , \theta)$ below, depends from $r$ and from the polar angle $\theta$, i.e. the coordinate $\theta$ should not be confused with the additional field $\vartheta$.}
\[
    \frac{\psi \dot{\phi}}{\sqrt{1 - \dot{\phi} ^{2}}} = C (r, \theta)
    ,
\]
where $C(r, \theta)$ is a time independent function of integration. Solving for $\psi$ and thanks to
(\ref{eq:phiphidot}) we can find
\[
    \psi
    =
    \frac{(M - f \vartheta) C (r, \theta)}{V \sqrt{1 - (M - f \vartheta) ^{2} / V ^{2}}}
    ,
\]
or, equivalently,
\[
    \Phi
    =
    - \sqrt{-g} + \frac{(M - f \vartheta) C (r, \theta)}{V \sqrt{1 - (M - f \vartheta) ^{2} / V ^{2}}}
    .
    \nonumber
\]
Since the new term $\vartheta \Phi$ in the action is $g _{\mu \nu}$ independent, the expression for
$\rho _{\Phi}$ is still given by (\ref{eq:rhoPhi}). On the other hand, in the total energy density, $\rho _{\mathrm{total}}$,
we have to add a contribution from $\vartheta$: we will temporarily denote this contribution $\rho _{\vartheta}$,
waiting for the following discussion about ${\mathcal{L}} _{(\vartheta)}$ before giving a more precise characterization.
Then
\begin{eqnarray}
    \rho _{\mathrm{total}}
    & = &
    \left\{
        - \sqrt{-g} + \frac{(M - f \vartheta) C (r, \theta)}{V \sqrt{1 - (M - f \vartheta) ^{2} / V ^{2}}}
    \right\}
    \times
    \nonumber \\
    & & \qquad
    \times
    \frac{V ^{2}}{(M - f \vartheta) \sqrt{-g}}
    \times
    \left(
        1 - \frac{(M f \vartheta) ^{2}}{V ^{2}}
    \right)
    +
    \frac{V ^{2}}{(M - f \vartheta) \sqrt{-g}}
    +
    \rho _{\vartheta}
    \nonumber \\
    & = &
    (M - f \vartheta)
    +
    \frac{C(r, \vartheta)}{\sqrt{-g}}
    \sqrt{1 - \frac{(M - f \vartheta)}{V ^{2}}}
    +
    \rho _{\vartheta}
    .
    \nonumber
\end{eqnarray}
Correspondingly the pressure can be determined as,
\[
    p _{\mathrm{total}} = - (M - f \vartheta) + \rho _{\vartheta} .
    \nonumber
\]
Notice that, even if $\vartheta$ did not have a potential originally, such a term is effectively generated by the
$(M - f \vartheta)$ term and by the $(M - f \vartheta) ^{2}$ term inside of the square root that appear in the
expression for $\rho _{\mathrm{total}}$. With this in mind, in what follows we will take $\rho _{\vartheta}$ as
being simply a kinetic term. We will now discuss, in more detail, the shape of this effective potential.

\section{Effective potential for the $\vartheta$ field and inverse quintessence for ghost $\vartheta$}

Once again, for the cosmological case we take $C (r, \vartheta)$ so that $\rho _{\mathrm{total}}$ is homogeneous,
which requires (for FRW spacetimes)
\[
    C (r, \vartheta) = \frac{K \sin \theta}{\sqrt{1 - k r ^{2}}}
\]
($K$ being again a positive constant); this produces an energy density
\[
    \rho _{\mathrm{total}}
    =
    \rho _{\vartheta} - V _{\mathrm{eff.}}
    ,
\]
where the effective potential is
\begin{equation}
    V _{\mathrm{eff.}} = - (M - f \vartheta) - \frac{K}{a ^{3}} \sqrt{1 - \frac{(M - f \vartheta) ^{2}}{V ^{2}}}
\label{eq:ghoVef}
\end{equation}
and the choice for sign in front of $V _{\mathrm{eff.}}$ will become clear after the following discussion.
Indeed, in this way we see that even if $\vartheta$ does not originally have a potential, one is dynamically generated
anyway. For $\rho _{\vartheta}$ we choose just a kinetic energy density, $\rho _{\vartheta} = \pm \dot{\vartheta} ^{2} / 2$,
and the minus sign means that we are dealing with a ghost field. Only for the ghost choice the sign of the effective potential
in (\ref{eq:ghoVef}) is the appropriare one. It is the ghost choice, which is, in fact, more
interesting, since then the effective potential (\ref{eq:ghoVef}) for $\vartheta$ has a minimum for each value of $a$.

As in the case studied in \cite{bib:23}, where an effective potential for both DE and DM depends on
some scalar and is  minimized, we also follow this procedure.
%In our case the effective potential is minimized at the
%value of the DE given by
%\begin{equation}
%    M - f \vartheta = \frac{V ^{2} a ^{3} / K}{\sqrt{1 - V ^{2}  a ^{6} / K ^{2}}}
%    .
%\label{eq:gho_de}
%\end{equation}

For the regular choice $\rho _{\vartheta} = \dot{\vartheta} ^{2} / 2$, $V _{\mathrm{eff.}}$ has the opposite sign as
%compared to the ghost choice (\ref{eq:ghoVef}) and (\ref{eq:gho_de}), which corresponds to a maximum, not a minimum,
compared to the ghost choice (\ref{eq:ghoVef}), which corresponds to a maximum, not a minimum,
as $a \to \infty$. Coming back to the choice in which $\vartheta$ behaves as a ghost field, this produces an ``inverse
quintessence scenario'' for the vacuum energy density; as opposed to standard quintessence \cite{bib:33}, where zero
vacuum energy is approached as $a \to \infty$, here the DE component ($M - f \vartheta)$ in eq.~(\ref{eq:ghoVef})
is zero as $a \to 0$ and approaches its maximum value as $a \to \infty$.

It is also interesting to notice some features in the minimization of the effective potential (\ref{eq:ghoVef}). In particular
\[
    \frac{d V _{\mathrm{eff.}}}{d \vartheta}
    =
    f
    -
    \frac{K}{a ^{3} V ^{2}}
    \frac{f ( M - f \vartheta )}{\sqrt{1 - (M - f \vartheta) ^{2} / V ^{2}}}
    ,
\]
so that, if $f \neq 0$ and\footnote{The additional factor $K / (a ^{3} V ^{2})$ is, naturally, never negative.} $M - f \vartheta > 0$,
the condition $d V _{\mathrm{eff.}} / (d \vartheta) = 0$ can be solved (after squaring) for $M - f \vartheta$, to obtain
\begin{equation}
    M - f \vartheta
    =
    \frac{a ^{3} V ^{2}}{K} \left( 1 + \frac{a ^{6} V ^{2}}{K ^{2}} \right) ^{-1/2}
    .
\label{eq:taceneden001}
\end{equation}
From the above we would like to evaluate $- \dot{\vartheta} ^{2} / 2$, the contribution of the tachyon to the total energy density
$\rho _{\mathrm{total}}$; this will make explicit that, both, as $a \to 0$ and as $a \to \infty$, the tachyon gives a small contribution
to the total energy density, which means that our approach appears to be self-consistent, at least in those limits. Indeed, noting that
the only quantity that depends from $t$ is the scale factor $a$, from (\ref{eq:taceneden001}) we obtain
\[
    - f \dot{\vartheta}
    =
    \frac{3 V ^{2}}{K a ^{6}} \left( \frac{1}{a ^{6}} + \frac{V ^{2}}{K ^{2}} \right) ^{-3/2} \left( \frac{\dot{a}}{a} \right)
    .
\]
Then the kinetic term of the ghost $\vartheta$ field is
\[
    - \frac{\dot{\vartheta} ^{2}}{2}
    =
    -
    \frac{9 V ^{4}}{2 K ^{2} f ^{2} a ^{12}} \left( \frac{1}{a ^{6}} + \frac{V ^{2}}{K ^{2}} \right) ^{-3} \left( \frac{\dot{a}}{a} \right) ^{2}
    .
\]
Since $( \dot{a} / a ) ^{2} \propto \rho _{\mathrm{total}}$, when we consider the limit $a \to 0$, we have that
\[
    - \frac{\dot{\vartheta} ^{2}}{2}
    \stackrel{a \to 0}{\approx}
    \mathrm{const.} \times \rho _{\mathrm{total}} \times a ^{6}
    ;
\]
moreover, in the limit $a \to + \infty$, we have
\[
    - \frac{\dot{\vartheta} ^{2}}{2}
    \stackrel{a \to +\infty}{\approx}
    \mathrm{const.} \times \frac{\rho _{\mathrm{total}}}{a ^{12}}
    .
\]
So, in both cases, the ghost contributes very little (vanishingly in the limit) to the total energy density.

\section{Discussion and conclusions}

In this paper we have seen that the use of non conventional kinetic terms for a scalar field $\phi$ in the form
$V (\phi) \sqrt{- \det(g _{\mu \nu} + \partial _{\mu} \phi \partial _{\nu} \phi)}$ can provide a unified picture
for DE and DM, when such terms are considered in the framework of a two measures approach.

DE and DM exchange can be introduced in a way similar to what we have considered before to
describe the role of a dynamical measure $\Phi$ as a way to model a dynamical tension in string and
branes~\cite{bib:16,bib:17,bib:18,bib:19,bib:20,bib:21,bib:22}.
For this an additional scalar field $\vartheta$ can be introduced, such that an
integration constant $M$, with the role of DE density, is replaced by $M - f \vartheta$ ($f$ being
some coupling constant). This procedure automatically induces a DE density, that, in fact, goes
as $M - f \vartheta$, and an effective dust energy density, that goes as
$K \sqrt{1 - (M - f \vartheta) ^{2} / V ^{2}} / a ^{3}$, when solving for constant $V$.

Another byproduct of our framework is an effective potential, induced for the field $\vartheta$, which is thus
responsible for the DE/DM dynamics. The sign of the effective potential for $\vartheta$ depends on the sign we
take for the kinetic term for $\vartheta$. Most interesting from the phenomenological point of view is the
choice of $\vartheta$ as a ghost field. Indeed, from the choice of $\vartheta$ as a ghost field an ``inverse
quintessence scenario'' for the vacuum energy is obtained, when the DE $M - f \vartheta$ is zero
as $a \to 0$ and approaches its maximum value $V$ as $a \to \infty$. In both limits, we have shown that the
contribution of the ghost field to the total energy density is negligible.

It is interesting to compare our results with those of other models for DE and DM that have
been considered in the recent literature. For example in \cite{bib:24} an increase of the DE density
is also realized, although as opposed to our case there such increase is unbounded. We remark, however, that
in our case the DE density starts from zero and then grows to a finite fixed value at large times and
that this effect is realized in a unified model of DE and DM. Another model that realizes DE/DM unification
and has been studied in detail is the Chaplygin gas model \cite{bib:34,bib:35}: in this model it is also
possible to derive a tachyon type Lagrangian \cite{bib:36}. DE/DM unification models are safe if a
sufficient fraction of initial density perturbations collapse into a gravitationally bound condensate
that can provide cold dark matter seeds for large-scale structure formation \cite{bib:37} and we expect
to be able to perform this check also in our model, for instance along the lines discussed in \cite{bib:36}.
This and other more phenomenologically related aspects will be considered in a future work.

\ack

SA would like to thank the Department of Physics of Ben Gurion University for hospitality and also acknowledge partial
support from the Microscopic Quantum Structure \& Dynamics of Spacetime group at the Max Planck Institute for Gravitational
Physics (Albert Einstein Institute) during the late stages of this work.
EIG would like to thank David Polarsky, Douglas Singleton and Nattapong Yongram for discussions on the subject of this work.

\appendix

\section{\label{app:dynten}Strings and branes with dynamical tension}

Here we would like to review the role of the alternative measure $\Phi$ in the context of strings and brane theories,
and then also how this can be used to construct models with varying string and brane tensions: this will be analogous
to what we did in four dimensional cosmology.

In the context of Polyakov formulation of string and branes, we use the world sheet metric $\gamma _{ab}$ and the standard
(Riemannian) volume element $\sqrt{- \gamma} d ^{D} \sigma$, where $\gamma = \det(\gamma _{ab})$ is the determinant of
$\gamma _{ab}$. A central feature of this volume element is \emph{reparametrization invariance}, but reparametrization
invariance can also be obtained if we define $p$ scalars $\varphi ^{a}$, $a = 1, 2, \dots , p$ and use as the volume
element
\[
    \Phi
    =
    \epsilon _{i _{1} \dots i _{p}} \epsilon ^{\mu _{1} \dots \mu _{p}}
    \partial _{\mu _{1}} \varphi ^{i _{1}} \dots \partial _{\mu _{p}} \varphi ^{i _{p}} ,
\]
where $\epsilon ^{\mu _{1} \dots \mu _{p}}$ and $\epsilon _{i _{1} \dots i _{p}}$ are the alternating symbols. With this
definition $\Phi$ transforms exactly as $\sqrt{- \gamma}$ under reparametrization transformations.

A straightforward use of the measure $\Phi$ in string theory is somewhat problematic however. Indeed, if in the Polyakov
action
\[
    S _{\mathrm{P}} [ X ^{\alpha} , \gamma _{cd} ]
    =
    - T \int d \sigma ^{0} d \sigma ^{1} \sqrt{- \gamma} \gamma ^{ab} g _{\mu \nu} \partial _{a} X ^{\mu} \partial _{b} X ^{\nu}
\]
we simply replace $\sqrt{- \gamma}$ by
\[
    \Phi = \epsilon ^{ab} \epsilon _{ij} \partial _{a} \varphi ^{i} \partial _{b} \varphi ^{j}
\]
we obtain the action
\begin{equation}
    S _{1} [ X ^{\alpha} , \gamma _{cd} , \varphi ^{k}]
    \sim
    - \int d \sigma ^{0} d \sigma ^{1} \Phi \gamma ^{ab} g _{\mu \nu} \partial _{a} X ^{\mu} \partial _{b} X ^{\nu} .
\label{eq:meamodstract}
\end{equation}
The above is not satisfactory since a variation with respect to $\gamma ^{ab}$ gives
\[
    \Phi \partial _{a} X ^{\mu} \partial _{b} X ^{\nu} g _{\mu \nu} = 0 ,
\]
which means that either $\Phi = 0$ or that the induced metric on the string vanishes.

To improve the situation we notice that the use of the measure $\Phi$ opens new possibilities for allowed contributions to the action.
Let us consider, for instance, the case when a contribution of the form $\sqrt{- \gamma} L$ is a total derivative: then after changing
the measure, it could certainly by that $\Phi L$ is \emph{not} a total derivative. This is exactly the situation if
\[
    L = \frac{\epsilon ^{ab}}{\sqrt{-\gamma}} F _{ab} , \quad \mathrm{where} \quad F _{ab} = \partial _{a} A _{b} - \partial _{b} A _{a} .
\]
So, if we consider the action
\begin{equation}
    S = S _{1} + S _{\mathrm{gauge}} ,
\label{eq:newstract}
\end{equation}
 where $S _{1}$ is given by (\ref{eq:meamodstract}) and
\[
    S _{\mathrm{gauge}} [ \gamma _{cd} , \varphi ^{k} , A _{a}] = \int d \sigma ^{0} d \sigma ^{1} \Phi \frac{\epsilon ^{ab}}{\sqrt{- \gamma}} F _{ab} ,
\]
we see that (\ref{eq:newstract}) is much more interesting. It is conformally invariant, provided $\varphi ^{k}$ are
transformed as
\[
    \varphi ^{i} \longrightarrow \varphi ^{\prime \, i} = \varphi ^{\prime \, i} ( \varphi ^{j} ) , \quad \Phi \longrightarrow J \Phi
\]
where $J = \det ( \partial \varphi ^{\prime i} / (\partial \varphi ^{j}) )$ and $\gamma _{ab}$ transforms as
\[
    \gamma _{ab} \longrightarrow \gamma ^{\prime} _{ab} = J \gamma _{ab} .
\]
The variation of the action with respect to $\varphi ^{k}$ gives
\[
    \epsilon ^{ab} \partial _{b} \varphi ^{k} \partial _{a}
    \left(
        - \gamma ^{cd} \partial _{c} X ^{\mu} \partial _{d} X ^{\nu} g _{\mu \nu}
        + \frac{\epsilon ^{cd}}{\sqrt{- \gamma}} F _{cd}
    \right)
    =
    0 .
\]
If $\det( \epsilon ^{ab} \partial _{b} \varphi ^{k}) \neq 0$, which is true if $\Phi \neq 0$, then the equation just above implies
\begin{equation}
    - \gamma ^{cd} \partial _{c} X ^{\mu} \partial _{d} X ^{\nu} g _{\mu \nu}
    + \frac{\epsilon ^{cd}}{\sqrt{- \gamma}} F _{cd} = M = \mathrm{const} .
\label{eq:firequmodact}
\end{equation}
Considering then the variation with respect to $\gamma ^{ab}$ we obtain
\begin{equation}
    - \Phi
    \left(
        \partial _{a} X ^{\mu} \partial _{b} X ^{\nu} g _{\mu \nu}
        -
        \frac{1}{2} \gamma _{ab} \frac{\epsilon ^{cd}}{\sqrt{- \gamma}} F _{cd}
    \right)
    =
    0 .
\label{eq:secequmodact}
\end{equation}
Solving for $\epsilon ^{cd} F _{cd} / \sqrt{- \gamma}$ from (\ref{eq:secequmodact}) and substituting in (\ref{eq:firequmodact})
we get
\[
    \partial _{a} X ^{\mu} \partial _{b} X ^{\nu} g _{\mu \nu}
    -
    \frac{1}{2} \gamma _{ab} \gamma ^{cd} \partial _{c} X ^{\mu} \partial _{d} X ^{\nu} g _{\mu \nu}
    -
    \frac{1}{2} \gamma _{ab} M
    =
    0 ;
\]
taking the trace of the above equation gives $M = 0$.

If we now look at the equation of motion obtained from the variation of the gauge field $A _{c}$, we obtain
\[
    \epsilon ^{cb} \partial _{b} \left( \frac{\Phi}{\sqrt{- \gamma}} \right) = 0 ,
\]
which can be integrated to obtain $\Phi = T \sqrt{- \gamma}$; the constant of integration $T$ has indeed the meaning of a string tension.

All the above can be straightforwardly generalized to branes. Indeed, the relevant action for a $p$-brane is
\begin{eqnarray}
    S = S _{p} + S _{p-\mathrm{gauge}}
    \nonumber \\
    S _{p} [ X ^{\alpha} , \gamma _{cd} , \varphi ^{k}] = - \int d ^{p+1} \sigma \Phi
            \gamma ^{ab} \partial _{a} X ^{\mu} \partial _{b} X ^{\nu} g _{\mu \nu}
    \nonumber \\
    S _{p-\mathrm{gauge}} [ \gamma _{cd} , \varphi ^{k} , A _{b _{1} \dots b _{p}}] = \int d ^{p+1} \sigma \Phi
            \frac{\epsilon ^{a _{1} \dots a _{p+1}}}{\sqrt{- \gamma}} \partial _{[ a _{1}} A _{a _{2} \dots a _{p+1}]} ,
    \nonumber
\end{eqnarray}
with $\Phi$ now defined in terms of $p+1$ scalars as
\[
    \Phi
    =
    \epsilon ^{a _{1} \dots a _{p+1}} \epsilon _{j _{1} \dots j _{p+1}}
    \partial _{a _{1}} \varphi ^{j _{1}} \dots \partial _{a _{p+1}} \varphi ^{j _{p+1}}
    .
\]
The variation with respect to the gauge field $A _{a _{1} \dots a _{p}}$ gives
\[
    \epsilon ^{a _{1} \dots a _{p}} \partial _{a _{1}} \left( \frac{\Phi}{\sqrt{- \gamma}} \right) = 0 ,
\]
which again means $\Phi = T \sqrt{- \gamma}$, where $T = \mathrm{const.}$ is then a dynamically generated brane tension.
The equation of motion obtained from the variation of the $\varphi ^{j}$ fields gives (for $\Phi \neq 0$)
\[
    - \gamma ^{ab} \partial _{a} X ^{\mu} \partial _{b} X ^{\nu} g _{\mu \nu}
    + \frac{\epsilon ^{a _{1} \dots a _{p+1}}}{\sqrt{- \gamma}} \partial _{[a _{1}} A _{a _{2} \dots a _{p+1}]} = M .
\]
Solving for the last term on the right hand side and considering also the equation obtained from the variation with
respect to $\gamma _{ab}$, one gets
\[
    \gamma _{ab} = - \frac{p - 1}{M} \partial _{a} X ^{\mu} \partial _{b} X ^{\nu} g _{\mu \nu} .
\]
If $M < 0$ (since $p > 1$), then by rescaling we obtain that $\gamma _{ab}$ is the induced metric on the brane.

We can now consider the more general case of the coupling of strings and branes to external sources. If we add to the action
of the brane a coupling to a world sheet current $j ^{a _{2} \dots a _{p+1}}$,
\[
    S _{\mathrm{current}} = \int d ^{p+1} \sigma A _{a _{2} \dots a _{p+1}} j ^{a _{2} \dots a _{p+1}} ,
\]
then the variation with respect to $A _{a _{2} \dots a _{p+1}}$ gives
\begin{equation}
    \epsilon ^{a _{1} \dots a _{p+1}} \partial _{a _{1}} \left( \frac{\Phi}{\sqrt{- \gamma}} \right) = j ^{a _{2} \dots a _{p+1}} ,
\label{eq:gaufiepbrequ}
\end{equation}
and what will be practically interesting for us in this work will be the case when a bulk scalar field $\phi$ induces the
current $j ^{a _{2} \dots a _{p+1}}$, as in the following ($q$ will be some coupling constant)
\[
    j ^{a _{2} \dots a _{p+1}} = q \partial _{\mu} \phi \frac{\partial X ^{\mu}}{\partial \sigma ^{a}} \epsilon ^{a a _{2} \dots a _{p+1}}
                               = q \partial _{a} \phi \epsilon ^{a a _{2} \dots a _{p+1}} .
\]
Then (\ref{eq:gaufiepbrequ}) can be integrated to obtain
\[
    \frac{\Phi}{\sqrt{- \gamma}} = q \phi + M ,
\]
which is the analogous to equation (\ref{eq:PhiMetMas}) in the cosmological case.

\end{document}